\documentclass[12pt]{article}
\usepackage[dvips]{color}
\usepackage{epsfig}
\usepackage{amsmath}
\usepackage{graphicx}
\def\Box{\hbox{$\rlap{$\sqcup$}\sqcap$}}
\begin{document}
\title{\bf Transition Rate and the Photoelectric effect in the Presence of a Minimal Length  }

\author{B. Khosropour \thanks{E-mail: $b _ -khosropour@kazerunsfu.ac.ir$}\hspace{1mm}\\
 {\small {\em  Department of Physics, Faculty of Sciences,
Salman Farsi University of Kazerun , Kazerun ,73175-457, Iran}}\\
\\}
\date{\small{}}
\maketitle
\begin{abstract}
In this work, according to generalized uncertainty principle (GUP)
and time- dependent perturbation theory, the transition rate in the
present of a minimal length based on the Kempf algebra is studied.
Also, we find the absorption cross section in the framework of GUP.
The modified photoelectric effect is investigated and we show that
the differential cross section of photoelectric effect in the
framework of GUP is related to the isotropic minimal length scale.
The upper bound on the isotropic minimal length is estimated.
\\

\noindent
\hspace{0.35cm}



\end{abstract}
\newpage

\section{Introduction}
In the last decade, different theories of quantum gravity such as
the string theory, loop quantum gravity, noncommutative geometry and
doubly special relativity are proposed for finding the unification
between the general theory of relativity and the standard model of
particle physics [1]. Although these theories are different in
concepts, all of these studies lead to unique belief which predicts
the existence of a measurable minimal length scale. An immediate
consequence of existence of a minimal length is that the Heisenberg
uncertainty principle is modified. Nowadays the modified uncertainty
principle is called generalized uncertainty principle (GUP) [2]. The
generalized uncertainty principle corresponding to the modified
Heisenberg algebra can be written as
\begin{equation}
\triangle X \triangle P\geq\frac{\hbar}{2}\left[1+\beta(\Delta
P)^{2}\right],
\end{equation}
where $\beta$ is a positive parameter [3,4] and also, Eq. (1) yields
a minimal measurable length $(\triangle X)_{min}=\hbar\sqrt{\beta}$.
At this time, many studies have been done to compute the corrections
of quantum mechanics in the GUP framework. These investigations seem
to modify mechanical Hamiltonians at atomic scales [5-10]. In the
recent years, a lot of papers have been devoted to the gravity and
reformulations of quantum field theory in the presence of a minimal
length scale [11-25]. Kempf and his collaborators have introduced
finite resolution of length can be obtained from the deformed
Heisenberg algebra [26-28]. The Kempf algebra in a D-dimensional
space is characterized by the following deformed commutation
relations
\begin{eqnarray}
\left[X^{i},P^{j}\right] &=&i\hbar[(1+\beta
\textbf{P}^{2})\delta^{ij}+\beta'P^{i}P^{j}], \\
\nonumber \left[X^{i},X^{j}\right] &=& i\hbar\
\frac{(2\beta-\beta')+(2\beta+\beta')\beta \textbf{P}^{2}}{1+\beta
\textbf{P}^{2}}(P^{i}X^{j}-P^{j}X^{i}), \\
\nonumber \left[P^{i},P^{j}\right] &=& 0,
\end{eqnarray}
where$i,j=1,2,...,D$ and $\beta ,\; \beta'$ are two positive
deformation parameters. In Eq. (2), $X^{i}$ and $P^{i}$ are position
and momentum operators in the GUP framework. According to Eq. (2),
we can easy to find the following an isotropic minimal length scale
\begin{equation}
(\Delta X^{i})_{min}=\hbar\sqrt{(D\beta+\beta') } , \quad\forall
i\in \{1,2, \cdots ,D\}.
\end{equation}
It should be mentioned that in a $D+1$-dimensional space- time the
following Lorentz-covariant deformed algebra have introduced by
Quesne and Tkachuk [29,30]
\begin{eqnarray}
\left[X^{\mu},P^{\nu}\right] &=&-i\hbar[(1-\beta P_{\rho}P^{\rho})g^{\mu\nu}-\beta'P^{\mu}P^{\nu}],
\\ \nonumber
\left[X^{\mu},X^{\nu}\right] &=& i\hbar\
\frac{2\beta-\beta'-(2\beta+\beta')\beta P_{\rho}P^{\rho} }{1-\beta
P_{\rho}P^{\rho}}(P^{\mu}X^{\nu}-P^{\nu}X^{\mu}), \\ \nonumber
\left[P^{\mu},P^{\nu}\right] &=& 0,
\end{eqnarray}
where $\mu,\; \nu, \; \rho=0,1,2,\cdots,D$ and
$g_{\mu\nu}=g^{\mu\nu}=diag(1,-1,-1,\cdots,-1)$. In the present
work, we study the transition rate and photoelectric effect in the
presence of a minimal length. For this purpose, we consider the
formalism of time-dependent perturbation theory to the interactions
of atomic electron with the modification of classical radiation
field. The paper is organized as follows: In Sec. 2, the Hamiltonian
of atomic electron is obtained in the presence of a minimal length.
From this modified Hamiltonian we find the modified transition rate.
Also, we investigate the absorption cross section in the presence of
a minimal length. In Sec. 3, the photoelectric effect in the
presence of a minimal length is studied. According to this study we
obtain the upper bound on the isotropic minimal length. Our
conclusions are presented in Sec. 4. We use SI units throughout this
paper.

\section{Transition Rate and Absorption Cross section in the Presence of a Minimal length}
The purpose of this section is finding the transition rate and an
absorption cross section in the presence of a minimal length based
on the Kempf algebra. So that we must to introduce the
representation of modified position and momentum operators which are
satisfied Kempf algebra in Eq. (2). In Ref. [31], Stetsko and
Tkachuk introduced the approximate representation fulfilling the
Kempf algebra in the first order over the deformation parameters
$\beta$ and $\beta'$
\begin{eqnarray}
X^{i} &=& x^{i}+ \frac{2\beta-\beta'}{4}(\textbf{
p}^{2}x^{i}+x^{i}\textbf{p}^{2}), \\ \nonumber P^{i} &=&
p^{i}(1+\frac{\beta'}{2}\textbf{p}^{2}),
\end{eqnarray}
where  the operators $x^{i}$ , $p^{i}$ satisfy the canonical
commutation relation and $\textbf{p}^{2}=\sum_{i=1}^{D}p^{i}p^{i}$.
It is interesting to note that in the special case of
$\beta'=2\beta$, the position operators commute in linear
approximation over the deformation parameter$\beta$, i.e.
$[X^{i},X^{j}] =0$. The following representations, which satisfy
Kempf algebra in the special case of $\beta'=2\beta$, was introduced
by Brau [32]
\begin{eqnarray}
X^{i} &=& x^{i}, \\
P^{i} &=& p^{i}(1+\beta \textbf{p}^{2} ).
\end{eqnarray}

\subsection{Interaction of Atomic electron with the Radiation Field in the Presence of a Minimal Length}
In this section, let us obtain the time-dependent perturbation
theory to the interactions of atomic electron with the classical
radiation field. Classical radiation field means that the electric
or magnetic field derivable from a classical radiation field [33].
The basic Hamiltonian, with $A^{2}$ vanished, is
\begin{eqnarray}
H=\frac{p^{2}}{2m_{e}}+e\phi(x)-\frac{e}{m_{e}c}\textbf{A}\cdot\textbf{p},
\end{eqnarray}
where in the coulomb gauge $(\nabla\cdot\textbf{A}=0)$,we have used
$\textbf{A}\cdot\textbf{p}=\textbf{p}\cdot\textbf{A}$. We work with
a monochromatic field of the plane wave for
\begin{eqnarray}
\textbf{A}=2A_{0}\hat{\epsilon}\cos(\frac{\omega}{c}\hat{n}\cdot\textbf{x}-\omega
t)=A_{0}\hat{\epsilon}[\exp(i\frac{\omega}{c}\hat{n}\cdot\textbf{x}-i\omega
t)+\exp(-i\frac{\omega}{c}\hat{n}\cdot\textbf{x}+i\omega t)],
\end{eqnarray}
where $\hat{\epsilon}$ and $\hat{n}$ are the polarization and
propagation direction. In Eq. (9), $\hat{\epsilon}$ is perpendicular
to the propagation direction $\hat{n}$ and from Eq. (8) treat
$-\frac{e}{m_{e}c}\textbf{A}\cdot\textbf{p}$ as time- dependent
potential. Now, for obtaining the Hamiltonian in the presence of a
minimal length, we must replace the usual position and momentum
operators with the modified position and momentum operators
according to Eqs. (6) and (7), we have
\begin{eqnarray}
H_{modified}=\frac{[(\textbf{p}-\frac{e}{c}\textbf{A})(1+\beta(\textbf{p}-\frac{e}{c}\textbf{A})^{2})]^{2}}{2m_{e}}+e\phi(x).
\end{eqnarray}
After neglecting terms of order $\beta^{2}$ and higher in Eq. (10),
the modified Hamiltonian can be obtained as follows
\begin{eqnarray}
H_{modified}&=&\frac{p^{2}}{2m_{e}}+e\phi(x)-\frac{e}{m_{e}c}\textbf{A}\cdot\textbf{p}+\frac{\beta}{m_{e}}p^{4}\\
\nonumber
&-&4\beta\frac{e}{m_{e}c}\textbf{A}\cdot\textbf{p}p^{2}+\frac{4\beta}{m_{e}}(\frac{e}{c})^{2}(\textbf{A}\cdot\textbf{p})^{2}.
\end{eqnarray}
From above equation we can consider the following modified time-
dependent potential
\begin{eqnarray}
{\cal
V}=-\frac{e}{m_{e}c}\textbf{A}\cdot\textbf{p}-4\beta\frac{e}{m_{e}c}\textbf{A}\cdot\textbf{p}p^{2}+\frac{4\beta}{m_{e}}(\frac{e}{c})^{2}(\textbf{A}\cdot\textbf{p})^{2}.
\end{eqnarray}
As we know that the $\exp(-i\omega t)$- term is responsible for
absorption, while the $\exp(i\omega t)$ - term is responsible for
stimulated emission. Now, let us treat the absorption case in the
presence of a minimal length. We have
\begin{eqnarray}
{\cal
V}=-\frac{eA_{0}}{m_{e}c}(\exp(i\frac{\omega}{c}\hat{n}\cdot\textbf{x})\hat{\epsilon}\cdot
\textbf{p})-4\beta\frac{eA_{0}}{m_{e}c}[\exp(i\frac{\omega}{c}\hat{n}\cdot\textbf{x})(\hat{\epsilon}\cdot
\textbf{p})p^{2}]+\frac{4\beta}{m_{e}}(\frac{eA_{0}}{c})^{2}[\exp(2i\frac{\omega}{c}\hat{n}\cdot\textbf{x})(\hat{\epsilon}\cdot
\textbf{p})^{2}],
\end{eqnarray}
and the modified transition rate is obtained as follows
\begin{eqnarray}
(w_{i\rightarrow n})_{modified}=(w_{i\rightarrow
n})_{0}+\beta^{2}(w_{i\rightarrow n})_{ML},
\end{eqnarray}
where
\begin{eqnarray*}
(w_{i\rightarrow
n})_{0}&=&\frac{2\pi}{\hbar}[(\frac{eA_{0}}{m_{e}c})^{2}|\langle n
|\exp(i\frac{\omega}{c}\hat{n}\cdot\textbf{x})(\hat{\epsilon}\cdot
\textbf{p})|i
\rangle|^{2}]\delta(E_{n}-E_{i}-\hbar\omega),
\end{eqnarray*}
and
\begin{eqnarray*}
 (w_{i\rightarrow
n})_{ML}&=&\frac{2\pi}{\hbar}[(\frac{4eA_{0}}{m_{e}c})^{2}|\langle n
|\exp(i\frac{\omega}{c}\hat{n}\cdot\textbf{x})(\hat{\epsilon}\cdot
\textbf{p})p^{2}|i \rangle|^{2}]\delta(E_{n}-E_{i}-\hbar\omega)\\
\nonumber&+&\frac{2\pi}{\hbar}[(\frac{4(eA_{0})^{2}}{m_{e}c^{2}})^{2}\langle
n
|\exp(2i\frac{\omega}{c}\hat{n}\cdot\textbf{x})(\hat{\epsilon}\cdot
\textbf{p})^{2}|i \rangle|^{2}]\delta(E_{n}-E_{i}-2\hbar\omega).
\end{eqnarray*}

\subsection{Absorption Cross Section in the Presence of a Minimal Length}
In this section we want to obtain the absorption cross section in
the framework of GUP. An absorption cross section has a definition
as follow
\begin{eqnarray}
\frac{(\frac{Energy}{unit time})absorbed\quad by\quad the\quad
atom(i\rightarrow n)}{Energy \quad flux \quad of \quad the\quad
radiation\quad field},
\end{eqnarray}
where the energy flux of the radiation field is given by
\begin{eqnarray}
I=\frac{c}{\pi}|\textbf{E}\times\textbf{B}|.
\end{eqnarray}
In this case we use the following definition for the electromagnetic
field
\begin{eqnarray}
\textbf{E}=-\frac{1}{c}\frac{\partial \textbf{A}}{\partial t} ,
\textbf{B}=\nabla\times\textbf{A}.
\end{eqnarray}
Let us obtain the modified energy flux of the radiation field. From
Eq. (17), first we must find the classical electromagnetic field in
the presence of a minimal length so that we write the
electromagnetic field in Eq. (17) by using the modified position and
momentum operators which are satisfied the Quesne-Tkachuk algebra,
that is
\begin{eqnarray}
x^{\mu}\longrightarrow  X^{\mu}&=&x^{\mu}, \\ \nonumber
\partial^{\mu}\longrightarrow\nabla^{\mu}&:=&(1+\beta\hbar^{2}\Box)\partial^{\mu},
\end{eqnarray}
where $\Box:=\partial_{\mu}\partial^{\mu}$ is the d'Alembertian
operator. If we substitute Eq. (18) into Eq. (17), we will obtain
the following modified electromagnetic field
\begin{eqnarray}
\textbf{E}_{ML}(\textbf{x},t)&=&-\frac{1}{c}(1+\beta \hbar^{2}\Box)\frac{\partial \textbf{A}}{\partial t}\\
\nonumber
\textbf{B}_{ML}(\textbf{x},t)&=&\nabla^{\mu}\times\textbf{A}.
\end{eqnarray}
After inserting Eq. (9) into Eq. (19), we have
\begin{eqnarray}
\textbf{E}_{ML}(\textbf{x},t)&=&-2kA_{0}(1-2\beta \hbar^{2}k^{2})\cos(\textbf{k}.\textbf{x}-\omega t+\theta),\\
\nonumber \textbf{B}_{ML}(\textbf{x},t)&=&-2(1-2\beta
\hbar^{2}k^{2})\textbf{k}\times A_{0}\hat{\epsilon}
\cos(\textbf{k}.\textbf{x}-\omega t+\theta),
\end{eqnarray}
where $\textbf{k}=\frac{\omega}{c}\hat{n}$. According to Eqs. (16)
and (20), we can easy obtain the energy flux in the presence of a
minimal length as follows
\begin{eqnarray}
I_{modified}=\frac{c}{\pi}|\textbf{E}_{ML}(\textbf{x},t)\times
\textbf{B}_{ML}(\textbf{x},t)|=I_{0}+\beta I_{ML},
\end{eqnarray}
where
\begin{eqnarray}
I_{0}&=&\frac{(\omega A_{0})^{2}}{2\pi c},\\ \nonumber
I_{ML}&=&4\frac{(\omega A_{0})^{2}}{2\pi c}(\frac{\hbar
\omega}{c})^{2}.
\end{eqnarray}
Now, from Eqs. (14), (15) and (22), we can obtain the modified
absorption cross section as follows
\begin{eqnarray}
(\sigma_{abs})_{modified}
=(\sigma_{abs})_{0}+\beta(\sigma_{abs})_{ML},
\end{eqnarray}
where

\begin{eqnarray}
(\sigma_{abs})_{0}&=&(\sigma_{abs})_{0}=\frac{(w_{i\rightarrow
n})_{0}}{I_{0}}=\frac{4\pi^{2}\hbar \alpha}{m_{e}^{2}\omega}|\langle
n|\exp(i\frac{\omega}{c}\hat{n}\cdot \textbf{x})\hat{\epsilon}\cdot
\textbf{p}| i\rangle|^{2} \delta(E_{n}-E_{i}-\hbar\omega),\\
\nonumber (\sigma_{abs})_{ML}&=&\frac{(w_{i\rightarrow
n})_{ML}}{I_{ML}}=\frac{(4\pi c)^{2}
\alpha}{m_{e}^{2}\omega}^{3}\hbar|\langle
n|\exp(i\frac{\omega}{c}\hat{n}\cdot \textbf{x})(\hat{\epsilon}\cdot
\textbf{p})p^{2}| i\rangle |^{2}\delta(E_{n}-E_{i}-\hbar\omega)\\
\nonumber
 &+&\frac{(4\pi A_{0}\alpha)^{2}
c}{m_{e}^{2}\omega}^{3}\hbar|\langle
n|\exp(2i\frac{\omega}{c}\hat{n}\cdot
\textbf{x})(\hat{\epsilon}\cdot \textbf{p})^{2}| i\rangle
|^{2}\delta(E_{n}-E_{i}-2\hbar\omega).
\end{eqnarray}
In the above equation $\alpha$ is $\frac{e^{2}}{\hbar c}$.

\section{Photoelectric Effect in the Presence of a Minimal Length Scale}
The photoelectric effect means that the ejection of an electron when
an atom is placed in the radiation field. The process of
photoelectric effect is considered to be the transition from an
atomic state to a continuum state [33]. According to previous
section, $| i\rangle$ is the ket for an atomic state, while $|
n\rangle$ can be taken to be a plane- wave state $| k_{f}\rangle$.
Now, we want to study the modified differential cross section for
the photoelectric effect by using our earlier formula for modified
absorption cross section. To find the number of states it is
convenient to use the box normalization convention for plane- wave
states. Assuming a plane- wave state normalized that means if we
integrate the square modulus of its wave function for a cubic box of
side $L$, we obtain unity. Also, the state is considered to satisfy
the periodic boundary condition with periodicity of the side of the
box. In the limit $L\rightarrow \infty$, the number of states is
reduced to the number of dots in three-dimensional lattice space. If
we consider the energy of the final- state plane wave
($E=\frac{\hbar^{2}k_{f}^{2}}{2m_{e}}$) and the periodic boundary
condition, we can easy find the following number of states in the
interval between $E$ and $E+dE$
\begin{eqnarray}
\rho(E)=(\frac{L}{2\pi})^{3}\frac{m_{e}k_{f}}{\hbar^{2}}dEd\Omega,
\end{eqnarray}
where $d\Omega$ is the solid angle element. According to Eqs. (24)
and (25) the modified differential cross section for the
photoelectric effect is obtained as follows
\begin{eqnarray}
(\frac{d\sigma}{d\Omega})_{modified}=(\frac{d\sigma}{d\Omega})_{0}+\beta(\frac{d\sigma}{d\Omega})_{ML},
\end{eqnarray}
where
\begin{eqnarray}
(\frac{d\sigma}{d\Omega})_{0}&=&[\frac{4\pi^{2}\hbar
\alpha}{m_{e}^{2}\omega}|\langle
k_{f}|\exp(i\frac{\omega}{c}\hat{n}\cdot
\textbf{x})\hat{\epsilon}\cdot \textbf{p}| i\rangle|^{2}(\frac{L}{2\pi})^{3}\frac{m_{e}}{\hbar^{2}}k_{f},\\
(\frac{d\sigma}{d\Omega})_{ML}&=&\frac{(4\pi
c)^{2}\alpha}{m_{e}^{2}\omega^{3}\hbar}|\langle
k_{f}|\exp(i\frac{\omega}{c}\hat{n}\cdot
\textbf{x})(\hat{\epsilon}\cdot \textbf{p})p^{2}| i\rangle|^{2}
\\ \nonumber
&+&\frac{(4\pi A_{0}\alpha )^{2}c}{m_{e}^{2}\omega^{3}}|\langle
k_{f}|\exp(2i\frac{\omega}{c}\hat{n}\cdot
\textbf{x})(\hat{\epsilon}\cdot \textbf{p})^{2}|
i\rangle|^{2}](\frac{L}{2\pi})^{3}\frac{m_{e}}{\hbar^{2}}k_{f}.
\end{eqnarray}
By considering the initial- state wave function is the ground- state
hydrogen atom wave function, Eqs. (27) and (28) become
\begin{eqnarray}
(\frac{d\sigma}{d\Omega})_{0}&=&(\frac{4\pi^{2}\hbar\alpha}{m_{e}^{2}\omega})\hat{\epsilon}\cdot
\int d^{3}x\frac{\exp(-i\textbf{k}_{f}\cdot\textbf{x})}{L^{\frac{3}{2}}}\exp(i\frac{\omega}{c}\hat{n}\cdot\textbf{x})(-i\hbar\nabla)[\exp(\frac{-Zr}{a_{0}})(\frac{Z}{a_{0}})^{\frac{3}{2}}],\\
(\frac{d\sigma}{d\Omega})_{ML}&=&(\frac{(4\pi c)^{2}
\alpha}{m_{e}^{2}\omega^{3}\hbar})\hat{\epsilon}\cdot \int d^{3}
x\frac{\exp(-i\textbf{k}_{f}\cdot\textbf{x})}{L^{\frac{3}{2}}}\exp(i\frac{\omega}{c}\hat{n}\cdot\textbf{x})(-i\hbar\nabla)(-\hbar^{2}\nabla^{2})
[\exp(\frac{-Zr}{a_{0}})\\
\nonumber&\times&(\frac{Z}{a_{0}})^{\frac{3}{2}}] + (\frac{(4\pi
\alpha A_{0})^{2} c}{m_{e}^{2}\omega^{3}}) \int d^{3}
x\frac{\exp(-i\textbf{k}_{f}\cdot\textbf{x})}{L^{\frac{3}{2}}}\exp(2i\frac{\omega}{c}\hat{n}\cdot\textbf{x})(\hat{\epsilon}\cdot(-i\hbar\nabla))^{2}[\exp(\frac{-Zr}{a_{0}})(\frac{Z}{a_{0}})^{\frac{3}{2}}],
\end{eqnarray}
where $a_{0}$ is the Bohr radius. After using the integrating by
parts and the perpendicular $\hat{n}$ to $\hat{\epsilon}$, we will
obtain the modified differential cross section for the photoelectric
effect as follows
\begin{eqnarray}
(\frac{d\sigma}{d\Omega})_{0}&=&32e^{2}k_{f}\frac{(\hat{\epsilon}\cdot
\textbf{k}_{f})^{2}}{m_{e}c\omega}(\frac{Z}{a_{0}})^{5}\frac{1}{[(\frac{Z}{a_{0}})^{2}+q^{2}]^{4}},
\\
(\frac{d\sigma}{d\Omega})_{ML}&=&(128\pi)e^{2}k_{f}\frac{(\hbar^{2}\hat{\epsilon}\cdot
\textbf{k}_{f})^{2}}{m_{e}c^{3}\omega^{3}}(\frac{Z}{a_{0}})^{5}\frac{1}{[(\frac{Z}{a_{0}})^{2}+q^{2}]^{4}}[(k_{f}c)^{4}+\frac{(\hat{\epsilon}\cdot
\textbf{k}_{f}))^{2}(cA_{0})^{2}}{a_{0}m_{e}}],
\end{eqnarray}
where $\textbf{q}=(\textbf{k}_{f}-\frac{\omega}{c}\hat{n})$ and
$(\hat{\epsilon}\cdot\textbf{k}_{f})^{2}=k_{f}^{2}\sin^{2}(\theta)\cos^{2}(\varphi)$.
If we consider the first term $(\frac{d\sigma}{d\Omega})_{0}$, the
usual differential cross section and the second term is
$(\frac{d\sigma}{d\Omega})_{modified}$ the relative modification of
differential cross section can be obtained as follows
\begin{eqnarray}
\frac{\Delta
(\frac{d\sigma}{d\Omega})}{(\frac{d\sigma}{d\Omega})_{0}}=(\hbar\sqrt{5\beta})\frac{8\pi}{c^{2}\omega^{2}}[(k_{f}c)^{4}+\frac{(\hat{\epsilon}\cdot
\textbf{k}_{f}))^{2}(cA_{0})^{2}}{a_{0}m_{e}}].
\end{eqnarray}
Another hand, if we substitute $\beta'=2\beta$ into Eq. (3), we will
obtain the following isotropic minimal length
\begin{eqnarray}
(\Delta
X^{i})_{min}=\sqrt{\frac{D+2}{2}}\hbar\sqrt{2\beta},\quad\forall
i\in \{1,2, \cdots ,D\}.
\end{eqnarray}
The isotropic minimal length in three spatial dimensions is given by
\begin{eqnarray}
(\Delta X^{i})_{min}=\hbar\sqrt{5\beta}.
\end{eqnarray}
Hence, by inserting Eq. (35) into Eq. (33), we will obtain the
relative differential cross section as follows
\begin{eqnarray}
\frac{\Delta
(\frac{d\sigma}{d\Omega})}{(\frac{d\sigma}{d\Omega})_{0}}=(\Delta
X^{i})_{min}^{2}\frac{8\pi}{c^{2}\omega^{2}}[(k_{f}c)^{4}+\frac{(\hat{\epsilon}\cdot
\textbf{k}_{f})^{2}(cA_{0})^{2}}{a_{0}m_{e}}].
\end{eqnarray}
Now we can estimate the upper bound on the isotropic minimal length
in modified photoelectric effect. If we consider the value of
differential cross section is about $10^{-28}m^{2}$ and also
considering $\omega\approx10^{14} Hz, a_{0}\approx10^{-11}m,
K_{f}\approx 10^{18}m^{-1}$. Here, according to Eq. (36), we can
estimate the following upper bound on the the isotropic minimal
length
\begin{eqnarray}
10^{-28}&\approx&(\Delta X^{i})_{min}^{2}10^{32},\\ \nonumber
(\Delta X^{i})_{min}&\leq&10^{-30}m.
\end{eqnarray}
It is interesting to note that the numerical value of the upper
bound on isotropic minimal length in Eq. (37), is close to the
Planck length scale($L_{p}\approx10^{-35}m$).

\section{Conclusions}
Heisenberg believed that every theory of elementary particles
contain a minimal length scale [34,35]. Today we know that every
theory of quantum gravity predicts a minimal observable distance. An
immediate consequence of GUP is a modification of position and
momentum operators. This study has found the transition rate and
photoelectric effect in the presence of a minimal length. First, we
have considered the time- dependent perturbation theory and then the
modified Hamiltonian of atomic electron was obtained up to the first
order over the deformation parameter $\beta$. According to the
modified Hamiltonian the transition rate in the framework of GUP was
investigated. We have seen that the modified transition rate was
included in two terms, one term was usual transition rate and the
second term was its correction due to the considered minimal length
effect. Hence, we have assumed the modified cross section in two
terms and then two terms of modified cross section have been
obtained. It is necessary to note that , in the limit
$\beta\rightarrow0$, the modified cross section become the usual
cross section. Also, the photoelectric effect in the presence of a
minimal length was investigated. We have shown that the relative
differential cross section was related to the isotropic minimal
length. The upper bound on the isotropic minimal length scale has
been estimated. It is interesting to note that the upper bound on
the isotropic minimal length was close to the Planck length scale.

%\section*{Acknowledgments}

\end{document}